\documentclass[reprint, aps, pre, notitlepage]{revtex4-1}

\usepackage{amsmath, amsfonts, amssymb, amsthm, mathtools}
\usepackage[utf8]{inputenc}
\usepackage[T1]{fontenc}
\usepackage{hyperref}
\usepackage{graphicx}
\usepackage[caption=false]{subfig}
\usepackage{lipsum}
\usepackage{xcolor}

\theoremstyle{remark}

\theoremstyle{remark}

\theoremstyle{theorem}
\newtheorem{thm}{Theorem}

\theoremstyle{theorem}

\newcommand{\hyp}{\,_2F_1}

\begin{document}
	
\title{Statistics of occupation times and connection to local properties of non-homogeneous random walks}
\author{Mattia Radice}
\email[Corresponding author: ]{m.radice1@uninsubria.it}
\author{Manuele Onofri}
\author{Roberto Artuso}
\author{Gaia Pozzoli}
\affiliation{Dipartimento di Scienza e Alta Tecnologia and Center for Nonlinear and Complex Systems, Università degli studi dell'Insubria, Via Valleggio 11, 22100 Como, Italy}
\affiliation{I.N.F.N. Sezione di Milano, Via Celoria 16, 20133 Milano, Italy}
\begin{abstract}
We consider the statistics of occupation times, the number of visits at the origin and the survival probability for a wide class of stochastic processes, which can be classified as renewal processes. We show that the distribution of these observables can be characterized by a single parameter, that is connected to a local property of the probability density function (PDF) of the process, viz., the probability of occupying the origin at time $ t $, $ P(t) $. We test our results for two different models of lattice random walks with spatially inhomogeneous transition probabilities, one of which of non-Markovian nature, and find good agreement with theory. We also show that the distributions depend only on the occupation probability of the origin by comparing them for the two systems: when $ P(t) $ shows the same long-time behavior, each observable follows indeed the same distribution.
\end{abstract}
	
	\maketitle
\section{Introduction}
Survival and persistence problems for stochastic processes are often considered in the study of critical phenomena in equilibrium and nonequilibrium systems, for instance spin systems in one or higher dimensions \cite{Stauffer-1994}, phase-ordering kinetics \cite{BraDerGod-1994, KraBenRed-1994, MajSire-1996} and twisted nematic liquid crystals exhibiting planar Ising model dynamics \cite{YurParMaj-1997}. Such problems are characterized by the \textit{persistence exponent} \cite{Bray-2000}, which gives the scaling of the probability that the order parameter $ x(t) $ (for example, the magnetization of a ferromagnet) of a system quenched from the disordered phase to its critical point has not changed sign in a time interval $ t $ following the quench \cite{MajBraCorSir-1996, OerCorBra-1997}. In this context the time evolution of the order parameter is treated as a stochastic process and other questions regarding the statistics of $ x(t) $ naturally arise: for instance, one can ask what is the fraction $ t^+ $ of time in which the process has assumed positive values \cite{God-Luc, BelBar-2005, KorBar-2011}, which is associated, e.g., with the mean magnetization. For many physical systems \cite{DorGod-1998, DroGod-1998, TorNewDas-1999} the distribution of the mean magnetization displays a U-shaped curve, reflecting the fact that, contrary to intuition, the order parameter is more likely to preserve its sign during the observation time. Interestingly, it is found that the exponent of the singularities at the outer values is closely related to the persistent exponent. This connection can be proved \cite{God-Luc} by considering $ x(t) $ as generated by a renewal process: starting from the initial state $ x(t_0) $, during the time evolution the process resets itself to the initial condition at random times $ t_i $, $ i=1,2,\dots $, such that the intervals $ \Delta t_i=t_i-t_{i-1} $ are independent and identically distributed random variables. In this setting it is also worth asking what is the number of renewals observed up to time $ t $, which is found to follow a Mittag-Leffler distribution of an adequate parameter. The value of the parameter and therefore the shape of the distribution depend on the scaling exponent of the probability density function of waiting times between renewals: for a distribution which decays asymptotically as $ F(\Delta t)\sim \Delta t^{-1-\theta} $, with $ 0\leq\theta<1 $, one obtains a Mittag-Leffler of parameter $ \theta $ \cite{ HeBurMetBar-2008,SchBarMet-2013,MetJeoCheBar-2014,Lei-Bar}.

The advantage of renewal theory is that it applies to a broad range of stochastic processes, including, for example, random walks with spatially inhomogeneous transition probabilities and correlations between steps. Note that non homogeneous diffusion has attracted recently attention in a variety of different contexts, see for example \cite{DenCorSch2004, KuhIhaHyy-2011, NisBer-2018, AltMorWal-preprint}. The major difficulty in applying renewal theory to general diffusion processes is that one has to determine the waiting-time distribution, which is often a difficult task to perform, especially when translational symmetry is broken or for walks of non-Markovian nature. In the language of random walks, for example, this corresponds to computing the probabilities of first return to the starting position, which can be analitically done only in few cases. In this paper we will show that it is possible to obtain the fraction of time spent in the positive axis, the number of renewals and the persistence exponent just by considering the probability $ P(t) $ that at time $ t $ the process is returned to the initial state:
\begin{equation}\label{key}
P(t)=\Pr\left\{x(t)=x(t_0)\right\}.
\end{equation}

The paper is organized as follows: in the next section we introduce the class of processes to which our results apply; then we present the known results regarding the occupation time of the positive axis (Sec. \ref{s:Lamp}), the number of returns at the origin (Sec. \ref{s:DK}) and the survival probability (Sec. \ref{s:Surv}), and discuss how to establish a connection between the three, starting from the probability of occupying the initial state; in Sec. \ref{s:Num} we describe the stochastic processes we have considered in our simulations and show the numerical results; finally in Sec. \ref{s:Concl} we draw our conclusions.

\section{The class of processes considered in the paper}
The class of stochastic processes for which the results of this paper apply is similar to that considered in a classic paper by Lamperti \cite{Lam}. This class regroups processes (not necessarily Markovian), whose time evolution is described by a discrete parameter $ n $, with the property that the states are divided into two sets, say $ A $ and $ B $, which communicate through the occurrence of a recurrent state $ x_0 $, assumed as the initial state. More precisely, denoting with $ x_n $ the state at time $ n $ of the stochastic process, starting from $ x_0$, we consider processes such that if $ x_{n-1}\in A $ and $ x_{n+1}\in B $ or vice versa, then $ x_n=x_0 $; moreover, the occupation of $ x_0 $ is a \textit{persistent recurrent event} \cite{Fell-Trans}, by which we mean that, calling $ F_n $ the probability that the process returns to $ x_0 $ for the first time after $ n $ steps, having started from $ x_0$, then
\begin{equation}\label{eq:Pers_recurr}
\sum_{n=1}^{\infty}F_n=1,
\end{equation}
i.e., the return to $ x_0 $ is certain. We will furthermore assume that the return to $ x_0 $ defines a renewal event, so that $ x_n $ can be treated as a renewal process.

In practice, in this paper we will consider one-dimensional random walks on the integer lattice, starting from $ x_0=0 $, with nearest-neighbor jumps (without specifying the rules followed by the jumps). We will call $ A(B) $ the set of positive(negative) integers, and will assume that the occupation of state $ x_0 $, i.e., the return to the origin, is a persistent recurrent event.

\section{Occupation time of the positive axis}\label{s:Lamp}
For the class of processes we are considering in this paper, the result stated in \cite{Lam} provides the distribution, as the number of steps tends to infinity, of the fraction of time spent in the positive axis, which we call the \textit{Lamperti distribution} $ \mathcal{G}_{\eta,\rho}(\xi) $. Such a distribution is defined through two parameters. The first parameter is
\begin{equation}\label{eq:eta_def}
\eta=\lim_{n\to\infty}\mathbb{E}\left(\frac{k_n}{n}\right),
\end{equation}
where $ k_n $ denotes the occupation time of set $ A $ up to step $ n $, using the convention that the occupation of the origin is counted or not according to whether the last other state occupied was in $ A $ \footnote{This is the same convention used in \cite{Lam}}. Clearly $ \eta  $ is equal to $ 1/2 $ if the process is symmetric with respect to $ A $ and $ B $. The second parameter is defined as the limit
\begin{equation}\label{eq:rho_def}
\rho=\lim_{z\to 1}\frac{(1-z)F'(z)}{1-F(z)},
\end{equation}
where $ F(z) $ denotes the generating function of the recurrence times of $ x_0 $, i.e., the first return probabilities $ F_n $:
\begin{equation}\label{key}
F(z)=\sum_{n=1}^{\infty}F_nz^n.
\end{equation}
We have the following \cite{Lam}:
\begin{thm}\label{thm:Lamp}
	Let $ x_n $ be the process described above. Then
	\begin{equation}\label{key}
	\lim_{n\to\infty}\mathrm{Pr}\lbrace k_n/n\leq u\rbrace\equiv \mathcal{G}_{\eta,\rho}(u)
	\end{equation}
	exists if and only if both limits $ 0\leq\eta\leq 1 $, Eq. \eqref{eq:eta_def}, and $ 0\leq\rho\leq 1 $, Eq. \eqref{eq:rho_def}, exist. In this case $ \mathcal{G}_{\eta,\rho}(u) $ is the distribution on $ [0,1] $ which, provided both $ \eta $ and $ \rho\neq 0,1 $, has the density:
	\begin{equation}\label{eq:Lamp_dens}
	\mathcal{G'}_{\eta,\rho}(u)=\mathcal{N}\frac{u^\rho(1-u)^{\rho-1}+u^{\rho-1}(1-u)^{\rho}}{a^2u^{2\rho}+2au^\rho(1-u)^\rho\cos(\pi\rho)+(1-u)^{2\rho}},
	\end{equation}
	where
	\begin{align}\label{key}
	a&=\frac{1-\eta}{\eta}\\
	\mathcal{N}&=\frac{a\sin(\pi\rho)}{\pi}.
	\end{align}
	For $ \eta=0,1 $ and $ 0<u<1 $, the distribution is:
	\begin{equation}\label{key}
	\mathcal{G}_{\eta,\rho}(u)=
	\begin{cases}
	1 & \text{for } \eta=0 \\
	0 & \text{for }\eta=1;
	\end{cases}
	\end{equation}
 	for $ \rho=1 $ we have:
	\begin{equation}\label{key}
	\mathcal{G}_{\eta,1}(u)=
	\begin{cases}
	0 & \text{for }u<\eta\\
	1 & \text{for }u\geq\eta,
	\end{cases}
	\end{equation}
	while for $ \rho=0 $, $ 0\leq u<1 $:
	\begin{equation}\label{key}
	\mathcal{G}_{\eta,0}(u)=1-\eta.
	\end{equation}
\end{thm}

An important observation is that the existence of the limit \eqref{eq:rho_def} is equivalent to a condition on the form the generating function $ F(z) $ must assume, namely (see \cite{Lam}):
\begin{equation}\label{eq:rho_def_F}
F(z)=1-\left(1-z\right)^\rho L\left(\frac{1}{1-z}\right),
\end{equation}
where $ L(x) $ is a \textit{slowly-varying} function, by which we mean a continuous function, positive for large enough $ x $, that for any $ y>0 $ satisfies
\begin{equation}\label{key}
\lim_{x\to \infty}\frac{L(yx)}{L(x)}=1.
\end{equation}

Eq. \eqref{eq:rho_def_F} suggests that the distribution of the occupation time can be determined by evaluating the analytical expression of $ F(z) $. As we have already observed, however, in general the computation of the first return probabilities is hard to perform. Nevertheless, since we are taking as $ x_0 $ the site $ j=0 $, one can use a well-known formula, valid for any renewal process, relating $ F(z) $ to the generating function $ P(z) $ of the probabilities of occupying the origin at time $ n $, $ P_n $, which reads \cite{Hug}
\begin{equation}\label{eq:F_vs_P}
F(z)=1-\frac{1}{P(z)},
\end{equation}
to recast condition \eqref{eq:rho_def_F} as:
\begin{equation}
P(z) =\frac{1}{\left(1-z\right)^{\rho}}H\left(\frac{1}{1-z}\right),\label{eq:P(z)_form}
\end{equation}
where $ H(x)=1/L(x) $ is a slowly-varying function. In particular, Eq. \eqref{eq:P(z)_form} shows that the parameter $ \rho $ of the Lamperti distribution appears as an exponent in the generating function $ P(z) $. 

A first consequence is that the parameter $ \rho $ can be computed by evaluating $ P_n $. In order to show this, we make use of the following tauberian theorem \cite{Fell-II}:
\begin{thm}
	Let $ g_n\geq 0 $ and suppose that
	\begin{equation}\label{key}
	\sum_{n=0}^{\infty}g_nz^n=G(z)
	\end{equation}
	converges for $ 0\leq z<1 $. Then
	\begin{multline}\label{key}
	G(z)\sim\frac{1}{(1-z)^{\gamma}}\mathcal{H}\left(\frac{1}{1-z}\right),\; z\to 1^-\iff\\ g_0+\dots+g_n\sim\frac{1}{\Gamma(\gamma+1)}n^{\gamma}\mathcal{H}(n),\; n\to\infty
	\end{multline}
	where $ \mathcal{H}(x) $ is a slowly-varying function and $\gamma\geq0$.\\
	Furthermore, if the sequence $ \{g_n\} $ is ultimately monotonic and $ \gamma>0 $, it also holds
	\begin{multline}\label{key}
	G(z)\sim\frac{1}{(1-z)^{\gamma}}\mathcal{H}\left(\frac{1}{1-z}\right),\; z\to 1^-\iff\\ g_n\sim\frac{1}{\Gamma(\gamma)}n^{\gamma-1}\mathcal{H}(n),\; n\to\infty.
	\end{multline}
\end{thm}
By using Eq. \eqref{eq:P(z)_form} and applying the theorem, one has that, for $ 0<\rho\leq 1 $, $ P_n $ decays as
\begin{equation}\label{eq:Pn_decay}
P_n\sim\frac{1}{\Gamma(\rho)}\frac{H(n)}{n^{1-\rho}},
\end{equation}
meaning that $ \rho $ is related to the exponent appearing in the long-time limit of the occupation probability of the origin.

We remark that this result connects the behavior of the process regarding the occupation time of the sets $ A $ and $ B $, which is a non-local property, to a local property. For instance, for a simple symmetric random walk it is known that $ P_n $ decays with the power-law $ P_n\sim n^{-1/2} $, which corresponds to $ \rho=\tfrac 12 $. In this case the distribution of the occupation time follows the \textit{first arcsin law} \cite{Fell-I}, which is recovered by Theorem \ref{thm:Lamp} in the case $ \rho=\eta=\tfrac 12 $. For $ 0<\rho<1 $ the probability of being at the origin has the asymptotic decay $ P_n\sim n^{-(1-\rho)} $, up to a factor given by the slowly-varying function, and the distribution of the occupation time is represented by U-shaped curves. From formula \eqref{eq:Lamp_dens} we see that the divergence of these curves at $ u=0 $ and $ u=1 $ is given exactly by the exponent $ 1-\rho $. The situation is different for $ \rho=1 $: we have $ P_n\sim H(n) $, hence $ P_n $ does not decay as a power law. Instead it must behave for large $ n $ as a (ultimately) decreasing slowly-varying function, converging to a constant. In this case the occupation time is split among the two sets, in such a way that the process spends a fraction $ \eta $ of time in $ A $ and the remaining in $ B $. The distribution of the fraction of time in $ A $ is therefore a Dirac delta function centered around $ u=\eta $ and we will refer to this as the \textit{ergodic} case \cite{Lam}. In the opposite case, $ \rho=0 $, regarding $ P_n $ we can only conclude that
\begin{equation}\label{key}
\sum_{m=0}^nP_m\sim H(n),
\end{equation}
where this time $ H(n) $ must be (ultimately) increasing. Since by using Eqs. \eqref{eq:Pers_recurr} and \eqref{eq:F_vs_P}, one can show that a necessary and sufficient condition for recurrence is the divergence of $ P(z) $ \cite{Hug}, we can say that $ H(n) $ must diverge, but we expect the divergence to be slow. In this sense, the case $ \rho=0 $ corresponds to a crossover for the occupation of $ x_0 $ between being or not a persistent recurrent event. This can be better understood by observing that the distribution of the occupation time has masses $ m_A=\eta $ on $ u=1 $ and $ m_B = 1-\eta $ on $ u=0 $, meaning that the process spends all the time either in $ A $, with probability $ \eta $, or in $ B $, with probability $ 1-\eta $.

\section{Number of visits at the origin}\label{s:DK}
The number of renewals for a random walk is closely related to the number of visits at the origin. Indeed, if the return defines a renewal event, $ N $ visits at the starting site for a walker correspond to $ N-1 $ renewals. The distribution of the occupation time of the origin can be obtained from a classic result by Darling and Kac \cite{Dar-Kac}, where they showed that the limiting distribution of the occupation time of a set of finite measure for a Markov process is the Mittag-Leffler distribution:
\begin{equation}\label{key}
\mathcal{M}_\nu(\xi)=\frac{1}{\nu\xi^{1+\frac 1\nu}}L_{\nu}\left(\frac{1}{\xi^{\frac 1\nu}}\right),
\end{equation}
where $ L_\nu (x) $ denotes the L\'evy one-sided density of parameter $ \nu $, defined through the inverse Laplace transform from $ p $ to $ x $: $ L_\nu(x)=\mathcal{L}^{-1}\left[ \exp\left(-p^\nu\right)\right] $; the parameter $\nu  $ depends on the process itself. For the sake of clarity, here we briefly state the result, limiting ourselves to the case of random walks on a lattice - we point out, however, that the result holds in a more general setting.

Let $ x_n $ be a random walk on the integer lattice. Consider the generating function of the probabilities $ P_n(j|j_0) $ of arriving at site $ j $ in $ n $ steps, having started from $ j_0 $:
\begin{equation}\label{key}
P_z(j|j_0)=\sum_{n=0}^{\infty}P_n(j|j_0)z^n.
\end{equation}
Let $ V(j) $ be an integrable, non-negative function and suppose there exists a function $ \pi(z) $, $ \pi(z)\to\infty $ as $ z\to 1^- $, and a positive constant $ c $, such that
\begin{equation}\label{key}
\lim_{z\to 1^-}\frac{1}{\pi(z)}\sum_{j}P_z(j|j_0)V(j)= c,
\end{equation}
the convergence being uniform in $ E=\{j_0|V(j_0)>0\} $. Then the following result holds \cite{Dar-Kac}:
\begin{thm}
	For some normalizing sequence $ \{u_n\} $ the limiting distribution of
	\begin{equation}\label{key}
	\frac{1}{u_n}\sum_{m=0}^{n}V(x_m)
	\end{equation}
	exists and it is non-singular if and only if, for some $ 0\leq \nu <1 $,
	\begin{equation}\label{eq:DK_slowly-var}
	\pi(z)=\frac{1}{\left(1-z\right)^{\nu}}\mathcal{H}\left(\frac{1}{1-z}\right),
	\end{equation}
	where $ \mathcal{H}(x) $ is a slowly-varying function. Moreover, if \eqref{eq:DK_slowly-var} is satisfied, $ u_n $ can be taken to be $ c\pi\left(1-\tfrac 1n\right) $ and the limiting distribution is the Mittag-Leffler distribution $ \mathcal{M_\nu}(\xi) $.
\end{thm}

We will use this result to find the distribution of the occupation time of the origin. In order to do so, we take $ V(j)=\delta_{j,0} $ so that
\begin{equation}\label{key}
\sum_{j}P_z(j|j_0)V(j)=\sum_{j}P_z(j|j_0)\delta_{j,0}=P_z(0|j_0).
\end{equation}
Now, since $ \delta_{j,0}>0 $ only for $ j=0 $, we have to prove the existence of $ \pi(z) $ of the desired form, Eq. \eqref{eq:DK_slowly-var}, such that
\begin{equation}\label{key}
\lim_{z\to 1^-}\frac{P_z(0|0)}{\pi(z)}= c.
\end{equation}
Now, by definition,
\begin{equation}\label{key}
P_z(0|0)\equiv P(z)
\end{equation}
and we know from the discussion made in section \ref{s:Lamp}, see Eq. \eqref{eq:P(z)_form}, that
\begin{equation}\label{key}
P(z)=\frac{1}{\left(1-z\right)^{\rho}}H\left(\frac{1}{1-z}\right),
\end{equation}
where $ H(x) $ is slowly-varying. Then, for any positive constant $ c $, we can take
\begin{equation}\label{key}
\pi(z)=\frac{1}{c\left(1-z\right)^{\rho}}H\left(\frac{1}{1-z}\right)
\end{equation}
and have
\begin{equation}\label{key}
\lim_{z\to 1^-} \frac{P(z)}{\pi(z)}=c.
\end{equation}
Hence, for the theorem stated above, the limiting distribution of the random variable
\begin{equation}\label{eq:T_n_def}
T_n\equiv\frac{1}{H(n)n^{\rho}}\sum_{m=0}^{n}\delta_{x_m,0},
\end{equation}
describing the occupation time of the origin, is the Mittag-Leffler distribution of parameter $ \rho $, for $ 0\leq\rho<1 $.

We point out that when $ \rho=0 $ the Mittag-Leffler distribution becomes the exponential distribution, while for $ \rho=\tfrac 12 $ we have an half-gaussian: $ \mathcal{M}_{1/2}(\xi)=\pi^{-1/2 }\exp\left(-\xi^2/4\right) $, which is the limiting distribution for the simple symmetric random walk \cite{Fell-I}. For $ \rho=1 $ one has a degenerate case, with the convergence:
\begin{equation}\label{key}
\frac{1}{H(n)n}\sum_{m=0}^{n}\delta_{x_m,0}\to 1
\end{equation}
in probability, which is a kind of weak ergodic theorem \cite{Dar-Kac}, as the long-time limit of $ H(n) $ gives in this case the probability of occupying the origin, which decays to a constant (see discussion in Sec. \ref{s:Lamp}). This means that the process possesses a stationary distribution and the value of such a distribution at $ j=0 $ corresponds to the ensemble average of $ V(j)$. Therefore we have the convergence of the time average of $ V(j)$ over a single trajectory to its ensemble average, so that the density of $ T_n $ converges to a  Dirac delta function centered around $ \xi_0=1 $.

In appendix \ref{ap:Tn} we show that in the long-time limit $ T_n $ is proportional to the number of visits at the origin up to step $ n $, which we denote as $ M_n $, rescaled for its mean value:
\begin{equation}\label{key}
T_n\sim\frac{1}{\Gamma(1+\rho)}\frac{M_n}{\langle M_n\rangle}.
\end{equation}
We may therefore conclude that the result states that the random variable
\begin{equation}\label{key}
\xi=\lim_{n\to\infty}\frac{1}{\Gamma(1+\rho)}\frac{M_n}{\langle M_n\rangle}
\end{equation}
follows a Mittag-Leffler distribution of parameter $ \rho $, for $ 0\leq\rho<1 $, and a degenerate Mittag-Leffler distribution for $ \rho=1 $, whose density is
\begin{equation}\label{key}
P(\xi)=\delta(\xi-1).
\end{equation}

We remark that the result also holds if $ x_n $ is not a Markov process, provided that the return to $ x_0 $ defines a renewal event, so that the transition can be characterized by a waiting-time distribution between renewals, i.e., by the distribution of the first returns times. This happens, for example, if $ x_n $ is symmetric with respect to the starting point. Indeed, we could obtain the same from renewal theory, see \cite{HeBurMetBar-2008, SchBarMet-2013, MetJeoCheBar-2014, Lei-Bar} and references therein. It is worth observing that the parameter characterizing the distribution of the occupation time of the origin must be the same parameter of the Lamperti distribution, which in our setting describes the occupation time of the positive(negative) axis for a symmetric process.
We also point out that a similar connection was proved in \cite{God-Luc} by using scaling arguments, and also in the context of infinite ergodic theory for deterministic systems \cite{ThaZwe-2006}.

\section{Decay of the survival probability}\label{s:Surv}
As we have seen, the Lamperti and the Mittag-Leffler distribution are closely related, all due to the particular form that the generating functions $ P(z) $ and $ F(z) $ must assume. We will show that this is also related to the asymptotic decay of the survival probability in the set $ A(B) $. We define the survival probability in a set for a random walk on the integers with nearest-neighbor jumps as the probability $ Q_n $ of never leaving the set up to step $ n $. If $ A $ is the set of positive integers, then, following the convention in \cite{Lam} on how to count the occupation time, we have
\begin{equation}\label{key}
Q_n=\mathrm{Pr}\lbrace x_1\geq 0,x_2\geq 0,\ldots,x_n\geq 0|x_0=0\rbrace,
\end{equation}
with $ Q_0=1 $. Such a quantity can be computed exactly for random walks with i.i.d. jumps drawn from a continuous distribution, by using a well-known combinatorial identity known as the Sparre-Andersen theorem \cite{Sparre}:
\begin{equation}\label{key}
Q(z)=\sum_{n=0}^{\infty}Q_nz^n=\exp\left[\sum_{n=1}^{\infty}\frac{z^n}{n}\mathrm{Pr}\lbrace x_n\geq 0\rbrace\right].
\end{equation}
For any symmetric jump distribution, one obtains the behavior $ Q_n\sim n^{-1/2} $ for large $ n $, independently of the distribution itself. It can be shown that such a decay also holds for walks with nearest-neighbor jumps, i.e., a particular case of non-continuous jump distribution, provided that the jumps are symmetric, independent and identically distributed \cite{GodMajSch-2017}. Therefore, in the paradigmatic case of the simple symmetric random walk on the integers, one finds that the value $ \tfrac 12 $ describes the power-law decay of the survival probability and gives the correct parameter describing both the Lamperti and the Mittag-Leffler distributions. However, no results for the survival probability are available if jumps are correlated or not identically distributed.

In our setting, we can obtain a relation between the survival probability $ Q_n $ and the \textit{persistence} probability \cite{God-Luc}, namely the probability $ U_n $ of not observing any return up to time $ n $, which can be computed as
\begin{equation}\label{eq:Un_def_maintext}
U_n=1-\sum_{m=0}^nF_m.
\end{equation}
In appendix \ref{ap:Q_vs_U} we show that the generating functions $ Q(z) $ and $ U(z) $ satisfy the relation:
\begin{equation}\label{eq:Q(z)_vs_U(z)}
Q(z)=\frac{1+U(z)}{1+\left(1-z\right)U(z)}
\end{equation}
and that this implies $ Q(z)\sim U(z) $ as $ z\to 1 $, for $ 0\leq\rho<1 $. This means that the survival and the persistence probabilities have the same behavior for large $ n $.

By using Eq. \eqref{eq:Un_def_maintext} and the condition $ U_0=1 $, we can compute the generating function  
\begin{equation}\label{key}
U(z)=\frac{1-F(z)}{1-z}
\end{equation}
and hence, by using equation \eqref{eq:rho_def_F}, we find that $ U(z) $ must be of the form
\begin{equation}\label{eq:U(z)_sv}
U(z)=\frac{1}{(1-z)^{1-\rho}}L\left(\frac {1}{1-z}\right),
\end{equation}
where $ L(x)=1/H(x) $ is a slowly-varying function, and $ H(x) $ is the same appearing in equation \eqref{eq:P(z)_form}. Since, as we already stated, for $ z\to 1 $ we have $ Q(z)\sim U(z) $, the use of the tauberian theorem implies that the survival probability $ Q_n $ decays as
\begin{equation}\label{eq:Q_n_decay}
Q_n\sim\frac 1{\Gamma(1-\rho)}\frac{n^{-\rho}}{H(n)}.
\end{equation}
Once again, the quantity of interest is characterized by the Lamperti parameter.  We remark that this result holds for the class of processes we are considering, therefore not only for walks with i.i.d. jumps. For $ \rho=1 $ the tauberian theorem only assures that as $ n\to\infty $
\begin{equation}\label{key}
\sum_{m=0}^n Q_m\sim \frac 1{H(n)}
\end{equation}
where $ H(n) $ is a (ultimately) decreasing slowly-varying function, hence the asymptotic relation \eqref{eq:Q_n_decay} is not valid in this regime. We recall that in this case $ P_n $ does not decay as a power law, see Sec. \ref{s:Lamp}.

\section{Numerical results}\label{s:Num}
In this section we present numerical results for two different classes of walks. The first class is the \textit{Gillis random walk} \cite{Gill-bias}, which is a random walk on the integer lattice, starting from $ x_0=0 $, with non-trivial jump probabilities: depending on the position of the walker, the probabilities of jumping from site $ j $ to site $ j' $ are given by the following rules:
\begin{equation}\label{key}
p\left(j',j\right) =\frac{1}{2}\left(1-\frac{\epsilon}{j}\right)\delta_{j',j+1}+\frac{1}{2}\left(1+\frac{\epsilon}{j}\right)\delta_{j',j-1}
\end{equation}
for $ j\neq 0 $, and
\begin{equation}
p\left(j',0\right) = \frac{1}{2}\delta_{j',1}+\frac{1}{2}\delta_{j',-1}
\end{equation}
for $ j=0 $, where $ -1<\epsilon<1 $ is a real parameter. For $ \epsilon >0 $ there is a bias towards the origin, while for $ \epsilon<0 $ there is a bias away from it, which in both cases decreases with the distance from the origin. It can be shown that the random walk is recurrent only for $ \epsilon\geq-\tfrac 12 $ \cite{Gill-bias,Hug}, so we will consider this range only. We point out that this is one of the few examples of non-homogeneous random walks for which exact analytical results are given. Indeed, Gillis proved that the generating function $ P(z) $ reads:
\begin{equation}\label{eq:Gillis_genfun}
P(z)=\frac{\hyp\left(\tfrac{1}{2}\epsilon+1,\tfrac{1}{2}\epsilon+\tfrac{1}{2};1;z^2\right)}{\hyp\left(\tfrac{1}{2}\epsilon,\tfrac{1}{2}\epsilon+\tfrac{1}{2};1;z^2\right)},
\end{equation}
where $ \hyp(a,b;c;z) $ is the Gaussian hypergeometric function \cite{Abr-Steg}.

The second class of random walks we want to consider is the {\it averaged L\'evy-Lorentz gas} (ALL), which was presented in \cite{ACOR} and \cite{ACOR-2} in two different versions. This model is closely related to the well-known L\'evy-Lorentz gas, quite extensively studied in the literature \cite{BarFleKla-2000, BurCanVez-2010, BiaCriLenLig-2016, VezBarBur-2019, BiaLenPen-2020, BurVez-preprint}. The ALL consists in a generalization of the correlated random walk \cite{RH}: a particle starts from $ x_0=0 $ choosing with equal probability the initial direction of motion and performing only nearest-neighbor jumps. After each jump the walker can reverse the direction of motion with probability $ \mathfrak{r} $, which, depending on the value of a real parameter $ \alpha $, can assume a non-trivial position-dependent behavior: in the first version $ \mathfrak{r} $ decays as a power-law with the distance from the origin:
\begin{equation}\label{key}
\mathfrak{r}(j)\sim |j|^{-(1-\alpha)},\quad 0<\alpha<1.
\end{equation}
In the second version, instead, it is the transmission probability $ \mathfrak{t}=1-\mathfrak{r} $, i.e, the probability of preserving the direction of motion, that decays with the same power-law. In both cases at $ j=0 $ the reflection probability is $ \mathfrak{r}(0)=\tfrac 12 $. It is possible to derive the long-time properties of the model by using appropriate continuum limits, which lead to the following diffusion equation for the evolution of the PDF of the process:
\begin{equation}\label{eq:ALL_diff}
\frac{\partial P(x,t)}{\partial t}=\frac{\partial}{\partial x}\left[D_\alpha(x)\frac{\partial P(x,t)}{\partial x}\right].
\end{equation}   
Here $ D_\alpha(x) $ is a position-dependent diffusion coefficient, whose behavior is related to the form of the reflection probability. Eq. \eqref{eq:ALL_diff} corresponds to a Langevin equation interpreted following H\"{a}nggi-Klimontovich (isothermal interpretation). For a discussion on how different interpretations (It\^{o}, Stratonovich or H\"{a}nggi-Klimontovich) affect the statistical properties of the system, see \cite{Lei-Bar}. Here we only point out that one can compute the solution \cite{Reg-Gro-Far,Lei-Bar} to get the asymptotic growth of the mean square displacement for both versions of the model: in the first case, $ \langle x_t^2\rangle \sim t^{\frac2{1+\alpha}} $, hence transport is superdiffusive and we will refer to this as the \textit{superdiffusive} version; in the second case, $ \langle x_t^2\rangle \sim t^{\frac 2{3-\alpha}} $, transport is subdiffusive and we will name this the \textit{subdiffusive } version. We can obtain in both cases the asymptotic behavior of the probability density function at $ x=0 $, which reads:
\begin{equation}\label{eq:ALL_P(t)}
P(0,t)\sim
\begin{cases}
t^{-1/(1+\alpha)} & \text{superdiffusive}\\
t^{-1/(3-\alpha)} & \text{subdiffusive}.
\end{cases}
\end{equation}

From the results in Eqs. \eqref{eq:Gillis_genfun} and \eqref{eq:ALL_P(t)} we are able to compute the parameter $ \rho $ for both models, yielding the shape of the Lamperti distribution, the form of the Mittag-Leffler distribution and the asymptotic decay of the survival probability.

\subsection{Evaluation of the Lamperti parameter}
In order to obtain the analytical predictions to compare with the simulations results, all we need to do is to evaluate the Lamperti parameter $ \rho $.

For the Gillis random walk we use the analytical result regarding the generating function, Eq. \eqref{eq:Gillis_genfun}.
By using the properties of the hypergeometric function \cite{Abr-Steg}, we can rewrite $ P(z) $ in the form given in Eq. \eqref{eq:P(z)_form}, obtaining (see appendix \ref{ap:rho_Gillis}):
\begin{equation}\label{key}
\rho =
\begin{cases}
0 & \text{for } \epsilon=-\tfrac 12 \\
\tfrac 12+\epsilon & \text{for }-\tfrac 12<\epsilon<\tfrac 12\\
1 &  \text{for }\tfrac 12 \leq\epsilon<1.
\end{cases}
\end{equation}

For the averaged L\'evy-Lorentz gas we evaluate $ \rho $ by using the long time asymptotics of the probability of occupying the origin, which can be computed performing a continuum limit. As we reported in Eq. \eqref{eq:ALL_P(t)}, it can be shown that the probability $ P_n $ decays as \cite{ACOR,ACOR-2}
\begin{equation}\label{eq:LL-rho}
P_n\sim
\begin{cases}
n^{-1/(1+\alpha)} & \text{superdiffusive}\\
n^{-1/(3-\alpha)} & \text{subdiffusive},
\end{cases}
\end{equation}
where $ 0<\alpha<1 $. Since the exponent is connected to $ \rho $, see Eq. \eqref{eq:Pn_decay}, we immediately get:
\begin{equation}\label{key}
\rho=
\begin{cases}
1-\tfrac 1{1+\alpha} & \text{superdiffusive}\\
1-\tfrac 1{3-\alpha} & \text{subdiffusive}.
\end{cases}
\end{equation}

\subsection{Occupation time of the positive axis}
Here we provide the results of simulations regarding the occupation time of the set $ A $ for the Gillis random walk and both versions of the averaged L\'evy-Lorentz gas.

\begin{figure}
\subfloat[]{
	\includegraphics[width=8.5cm]{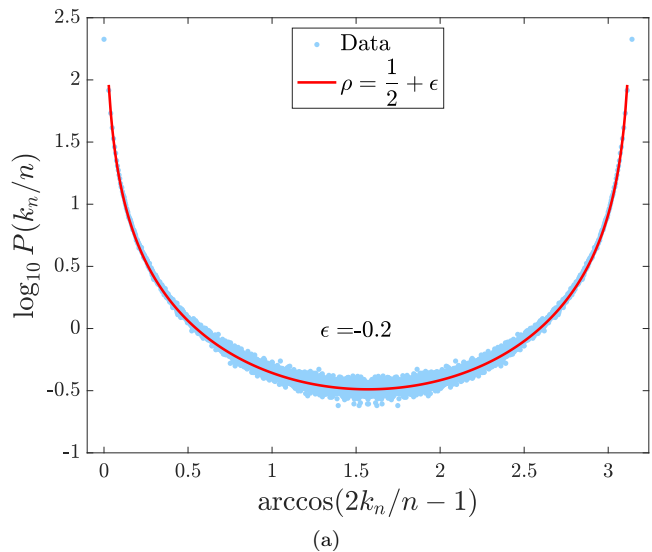}%
}\vfill
\subfloat[]{
	\includegraphics[width=8.5cm]{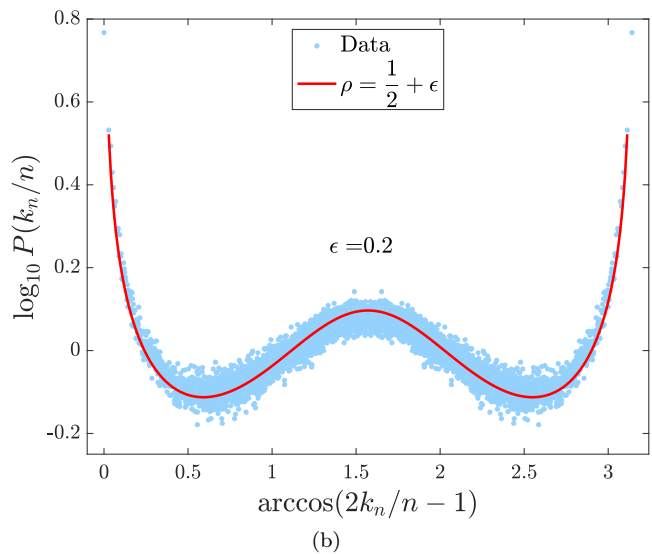}%
}
\caption{Distribution of the fraction of time spent in the positive axis for the Gillis random walk, in semi-logarithmic scale. To enhance readability of the outer values, it has been performed the transformation $ x\to \arccos(2x-1) $ on the $ x $-axis. The (light blue) dots represent the simulations results, the (red) line the theoretical prediction. The first plot displays the case $ \epsilon=-0.2 $, the second one $ \epsilon=0.2 $. In both cases the results are obtained simulating $ 10^6 $ walks of $ 10^4 $ steps.}
\label{fig:Gillis_Lamp}
\end{figure}

For the Gillis random walk we can recognize two different behaviors. When $ \epsilon<0 $ there is a bias away from the origin, and the distribution of the occupation time is represented by a U-shaped curve, meaning that the particle most likely spends all the time in one of the two sets. When $ \epsilon>0 $ there is a bias towards the origin, so we expect an higher contribution from walks which spend an equal amount of time in the two sets. This is confirmed by the plots in Fig. \ref{fig:Gillis_Lamp}, where we consider the cases $ \epsilon=-0.2 $ and $ \epsilon=0.2 $.

\begin{figure}
\includegraphics[width=8.5cm]{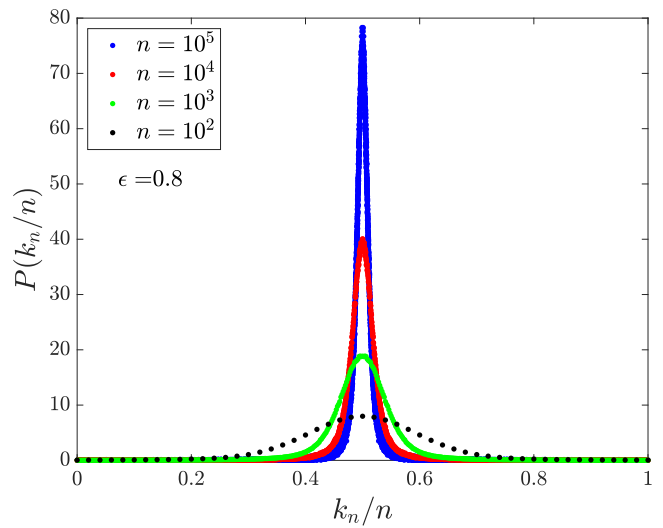}
\caption{Distribution of the fraction of time spent in the positive axis for the Gillis random walk, ergodic case, with $ \epsilon=0.8 $. Data are obtained simulating $ 10^6 $ walks of different numbers of steps. As the maximum number of steps grows, the distribution converges to a Dirac delta, centered around $ k_n/n=1/2 $.}
\label{fig:Gillis_Lamp_ergodic}
\end{figure}

When the bias towards the origin becomes sufficiently strong, i.e., for $ \epsilon\geq \tfrac 12 $, the outer values of the distribution cease to be the most probable. The process enters in an ergodic regime where the fraction of time spent in a set converges to its expected value, which in our case is $ \eta=\tfrac 12 $. In other words, the distribution is a Dirac delta function centered around $ \eta $. Fig. \ref{fig:Gillis_Lamp_ergodic} shows the behavior of the distribution as the number of steps grows, for $ \epsilon=0.8 $, confirming the convergence to a Dirac delta distribution.

\begin{figure}
	\subfloat{
		\includegraphics[width=8.5cm]{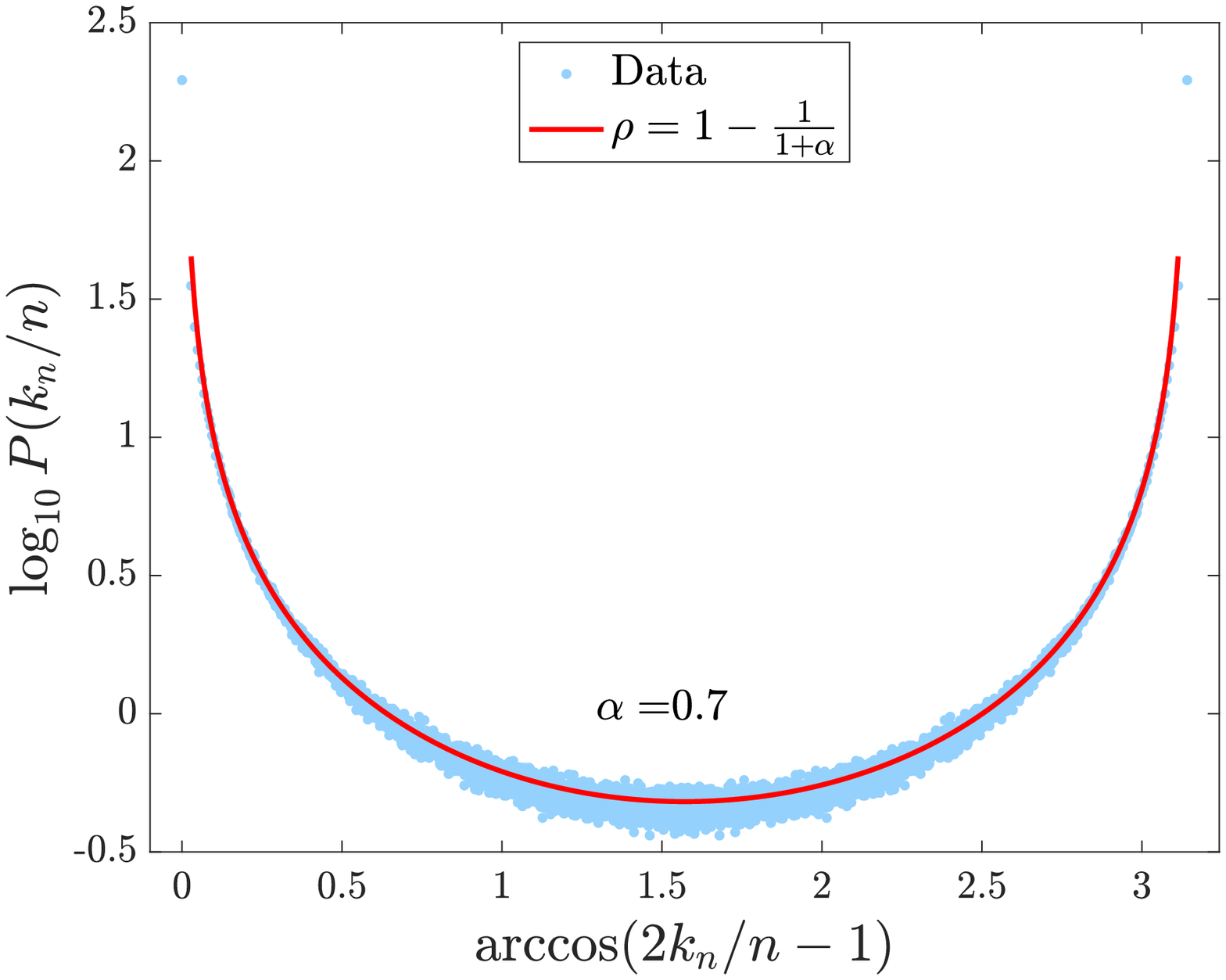}%
	}\vfill
	\subfloat{
		\includegraphics[width=8.5cm]{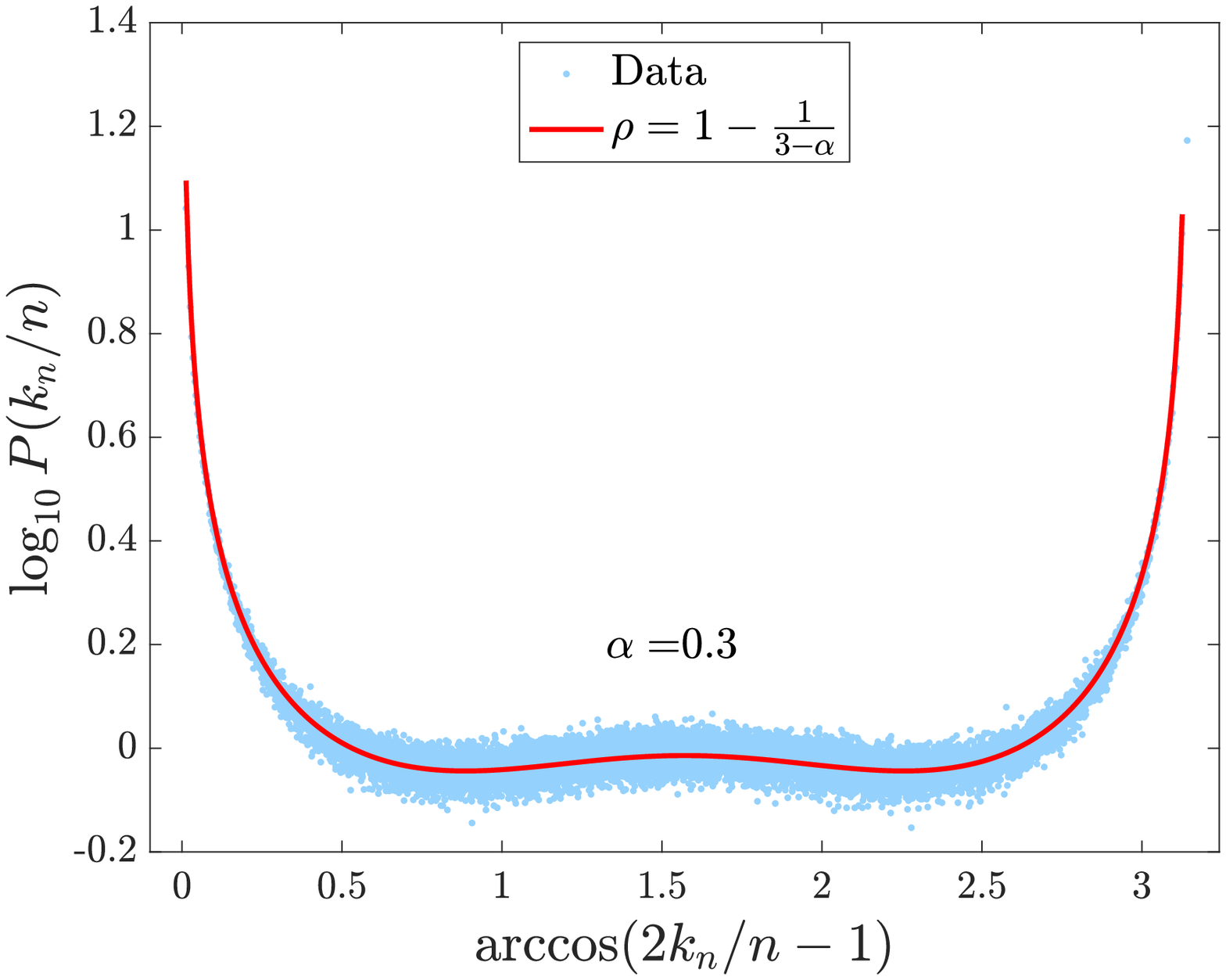}%
	}
	\caption{Distribution of the fraction of time spent in the positive axis for the averaged L\'evy-Lorentz gas, in semi-logarithmic scale. To enhance readability of the outer values, it has been performed the transformation $ x\to \arccos(2x-1) $ on the $ x $-axis. The (light blue) dots represent the simulations results, the (red) line the theoretical prediction. The first plot displays the superdiffusive case, with $ \alpha=0.7 $. Data are obtained simulating $ 10^6 $ walks of $ 10^4 $ steps. The second plot displays the subdiffusive version, with $ \alpha=0.3 $. In this case data are obtained simulating $ 10^7 $ walks of $ 10^5 $ steps.}
	\label{fig:LL_Lamp}
\end{figure}

For the averaged L\'evy-Lorentz gas, the behavior of the distribution depends on which version of the model we are considering. In the superdiffusive case the reflection probability decays as a power-law with the distance from the origin, $ \mathfrak{r}(j)\sim |j|^{-(1-\alpha)} $, therefore a particle tends to preserve its direction of motion as the distance from the starting point increases. As $ \alpha $ varies in $ (0,1) $, the Lamperti parameter varies in $ \left(0,\tfrac 12\right) $, see Eq. \eqref{eq:LL-rho}. In the subdiffusive case instead the reflection probability converges to $ 1 $ as the distance from the origin increases, with the transmission coefficient decaying as a power-law. The Lamperti parameter is in the range $ \tfrac 12<\rho<\tfrac 23 $. The behavior of both models is presented in Fig. \ref{fig:LL_Lamp}, for $ \alpha=0.7 $ in the superdiffusive case and $ \alpha=0.3 $ in the subdiffusive one. We point out that for both versions of the model we never enter the ergodic regime, as $ \rho\neq 1 $ for any value of $ \alpha $.

\subsection{Occupation time of the origin}
As we have already shown, the distribution of the occupation time of the origin follows a Mittag-Leffler distribution of the same parameter characterizing the Lamperti distribution. We consider the random variable
\begin{equation}\label{key}
\xi = \lim_{n\to\infty}\frac{1}{\Gamma(1+\rho)}\frac{M_n}{\langle M_n\rangle}.
\end{equation}

\begin{figure}
	\subfloat{
		\includegraphics[width=8.5cm]{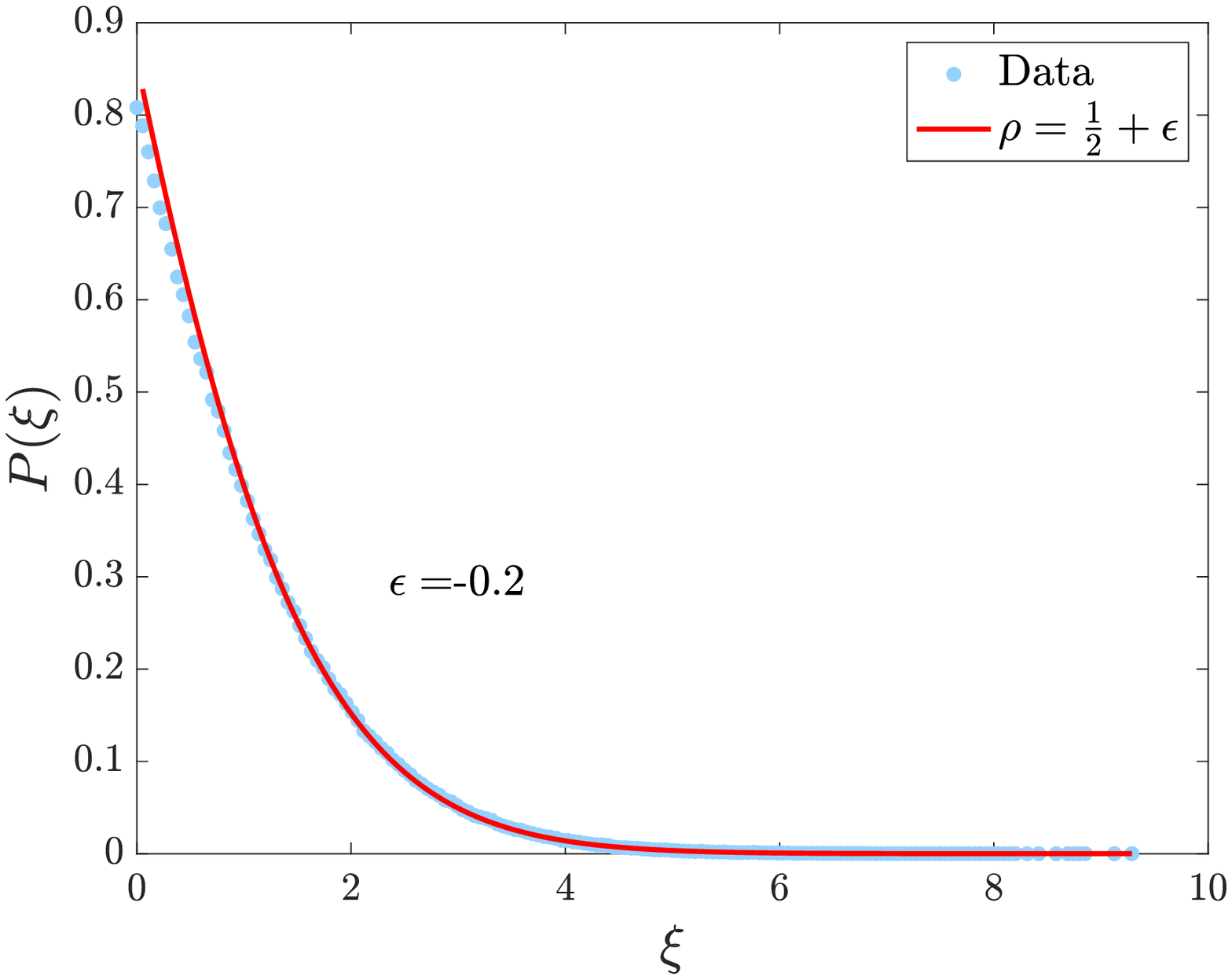}%
	}\vfill
	\subfloat{
		\includegraphics[width=8.5cm]{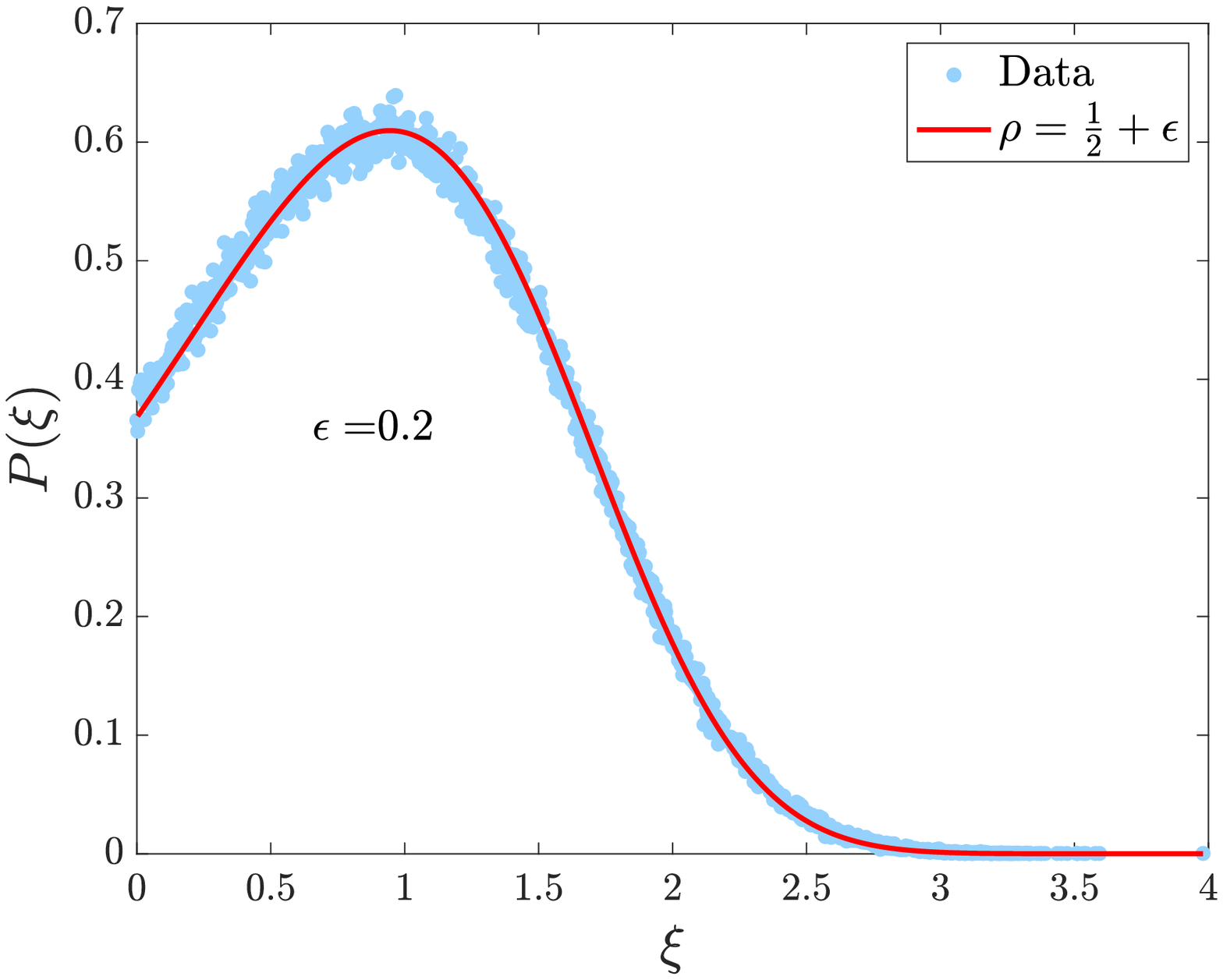}%
	}
	\caption{Distribution of the random variable $ \xi $ representing the rescaled number of steps in which the process occupies the origin. The (light blue) dots represent the simulations results, the (red) line the theoretical prediction. The first plot displays the case $ \epsilon=-0.2 $, the second one $ \epsilon=0.2 $. In both cases the results are obtained simulating $ 10^6 $ walks of $ 10^4 $ steps.}
	\label{fig:Gillis_ML}
\end{figure}

\begin{figure}
	\includegraphics[width=8.5cm]{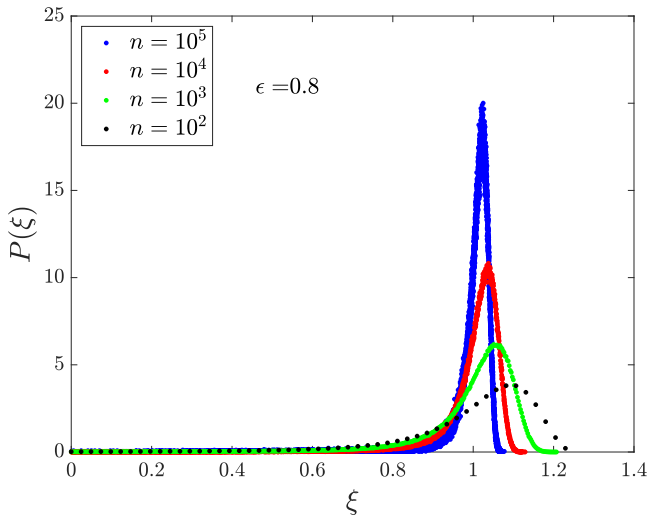}
	\caption{Distribution of the random variable $ \xi $ representing the rescaled number of steps in which the process occupies the origin, ergodic case, with $ \epsilon=0.8 $. Data are obtained simulating $ 10^6 $ walks of different numbers of steps. As the maximum number of steps grows, the distribution converges to a Dirac delta, centered around $ \xi_0=1 $.}
	\label{fig:Gillis_ML_ergodic}
\end{figure}

Once again the Gillis random walk is the model displaying the richest behavior.  As shown in Fig. \ref{fig:Gillis_ML}, for $ \epsilon<0 $ the distribution of $ \xi $ is monotonically decreasing, reflecting the fact that the particle is biased away from the origin. Indeed, the walks that do not return to the starting point have the highest probability. For $ \epsilon>0 $, instead, the bias is towards the origin, therefore the probability of returning increases. The shape of the distribution is quite different, and we have a pronounced peak close to $ \xi_0=1 $. For values $ \epsilon\geq\frac 12 $ we enter in the ergodic regime and the distribution converges to a Dirac delta function centered around $ \xi_0=1 $, Fig. \ref{fig:Gillis_ML_ergodic}.

\begin{figure}
	\subfloat{
		\includegraphics[width=8.5cm]{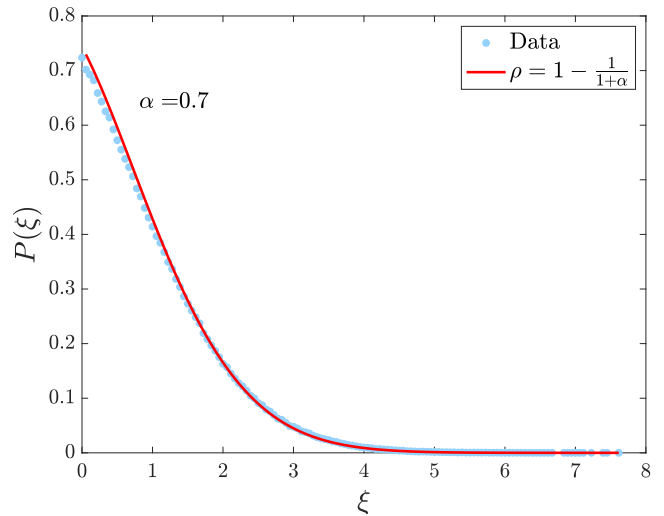}%
	}\vfill
	\subfloat{
		\includegraphics[width=8.5cm]{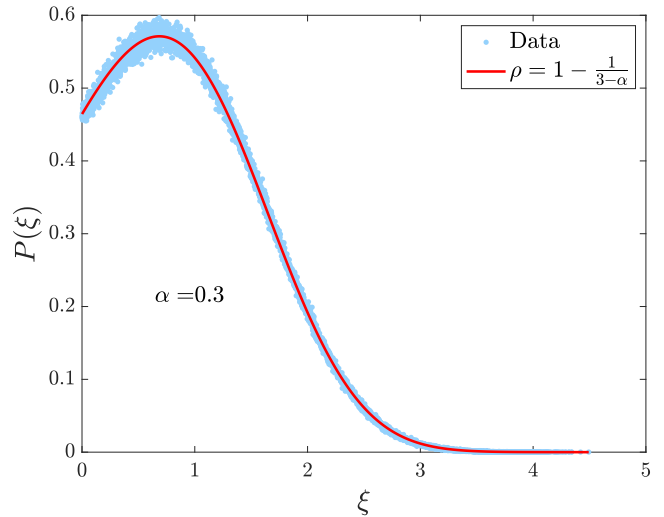}%
	}
	\caption{Distribution of the random variable $ \xi $ representing the rescaled number of steps in which the process occupies the origin. The (light blue) dots represent the simulations results, the (red) line the theoretical prediction. The first plot displays the superdiffusive case, with $ \alpha=0.7 $. The results are obtained simulating $ 10^6 $ walks of $ 10^4 $ steps. The second plot corresponds to the subdiffusive version, with $ \alpha=0.3 $. Data are obtained simulating $ 10^7 $ walks of $ 10^5 $ steps.}
	\label{fig:LL_ML}
\end{figure}

We can recognize a similar behavior for the averaged L\'evy-Lorentz gas, as shown in Fig. \ref{fig:LL_ML}. Here the shape of the distribution depends on which version of the model is considered: for the superdiffusive case we have a monotonically decreasing curve, while for the subdiffusive one the distribution presents a peak close to $ \xi_0=1 $. The values of $ \alpha $ chosen for the two systems are $ \alpha=0.7 $ for the superdiffusive version and $ \alpha=0.3 $ for the subdiffusive one.

\subsection{Decay of the survival and persistence probabilities}
For both models we also provide simulations regarding the asymptotic decay of the survival and persistence probabilities. As we have seen, we expect both quantities to decay as $ n^{-\rho} $, where $ \rho $ depends on a parameter characterizing the model ($ \epsilon $ for Gillis, $ \alpha $ for L\'evy-Lorentz). We confirm our prediction by plotting the exponent of the asymptotic decay of $ Q_n $ and $ U_n $, obtained from simulations, versus the characteristic parameter.

\begin{figure}
	\includegraphics[width=8.5cm]{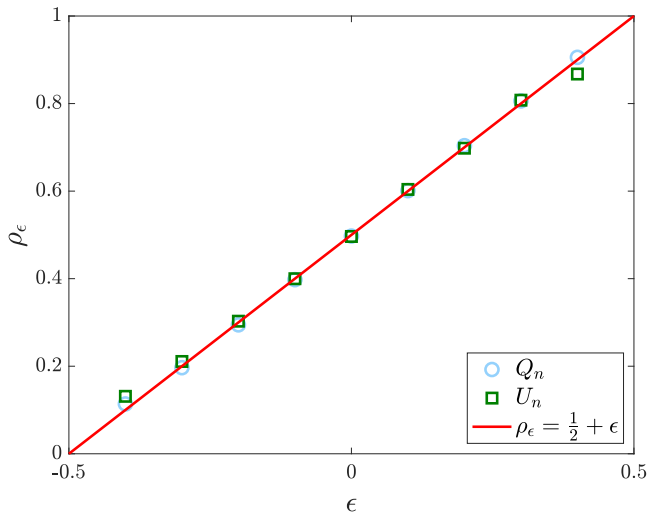}
	\caption{Exponents of the asymptotic power-law decay of the persistence and survival probabilities for the Gillis random walk. Data are obtained simulating $ 10^7 $ walks of $ 10^5 $ steps. The (green) squares represent the persistence probability, while the (light blue) circles refer to the survival probability.}
	\label{fig:Gillis_surv}
\end{figure}

For the Gillis random walk we have good agreement between the two computed exponents and the theoretical values, Fig. \ref{fig:Gillis_surv}. We point out that $ \epsilon $ is taken in the range $ \left(-\tfrac 12,\tfrac 12\right) $, so that $ 0<\rho<1 $. We observe that the agreement gets worse when $ \epsilon $ gets closer to the boundaries of the considered interval: we can explain this fact by considering that as $ \epsilon\to -\tfrac 12 $ convergence to the theoretical values becomes slower, while in the opposite case, $ \epsilon\to \frac 12 $, the system is getting closer to the regime $ \rho=1 $, where $ Q_n $ and $ U_n $ are not guaranteed to decay in the same way. 

\begin{figure}
	\subfloat{
		\includegraphics[width=8.5cm]{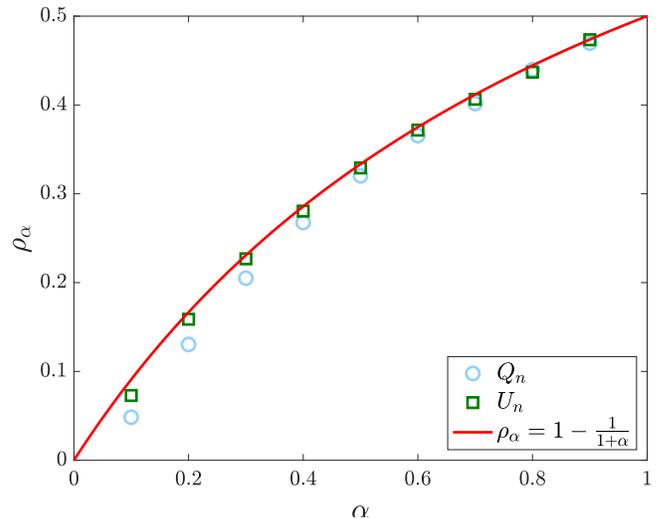}%
	}\vfill
	\subfloat{
		\includegraphics[width=8.5cm]{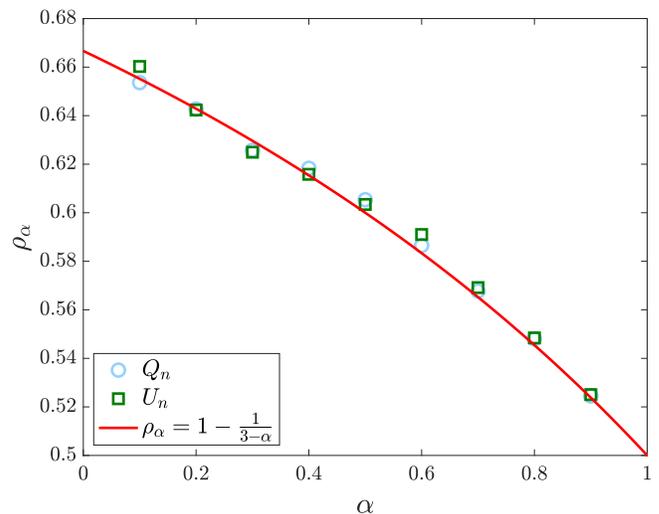}%
	}
	\caption{Exponents of the asymptotic power-law decay of the persistence and survival probabilities for the averaged L\'evy-Lorentz gas. The first plot displays the superdiffusive version, the second the subdiffusive one. In both cases data are obtained simulating $ 10^7 $ walks of $ 10^5 $ steps. The (green) squares represent the persistence probability, while the (light blue) circles refer to the survival probability.}
	\label{fig:LL_surv}
\end{figure}

For the superdiffusive averaged L\'evy-Lorentz gas we have good agreement when $ \alpha\geq 0.4 $, while for lower values of the parameter we observe a non-negligible difference between the two computed exponents. However, we point out that this is due to the fact that the continuum limit used to describe the long time properties of the system becomes effective after a preasymptotic regime, which depends on $ \alpha $, and the diffusive asymptotic regime is not yet captured at the number of steps of our simulations. Indeed, we observed in \cite{ACOR} the same discrepancies, in the same range of $ \alpha $, in the evaluation of the moments. For the subdiffusive version instead the difficulties to capture the asymptotic regime may be traced back to the fact that in order to observe cleanly the decay of the quantities of interest we need a larger number of steps with respect to the superdiffusive version. However, for both versions of the model we have in general a good agreement with the theoretical predictions, Fig. \ref{fig:LL_surv}.

\subsection{Comparison of different systems with the same Lamperti parameter}
From the discussion made so far it should be clear that the Lamperti parameter $ \rho $, characterizing the distributions of the observables we have considered in this paper, only depends on a local property of the PDF of the process, namely the probability $ P_n $ of occupying the origin at time $ n $. It can happen that two stochastic processes are described by two different sets of evolution laws, but share the same asymptotic power-law decay for the distribution of the occupation time of the origin, i.e., the $ P_n $ decay with the same exponent. As a consequence, the distributions of the occupation time of the positive axis and the number of returns to the origin will be the same.

\begin{figure}
	\subfloat{
		\includegraphics[width=8.5cm]{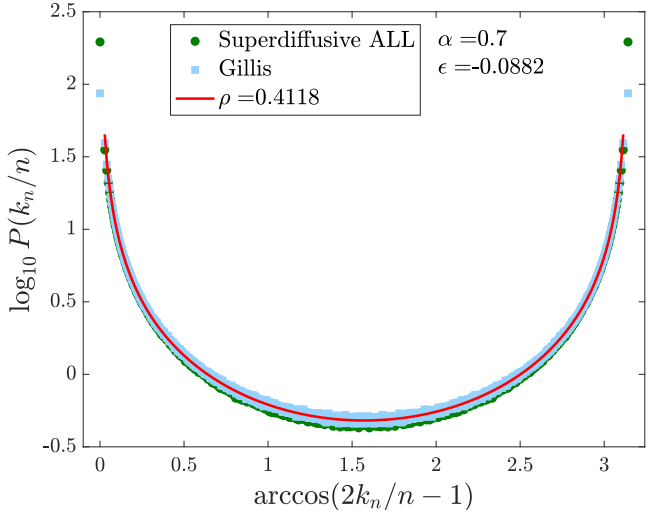}%
	}\hfill
	\subfloat{
		\includegraphics[width=8.5cm]{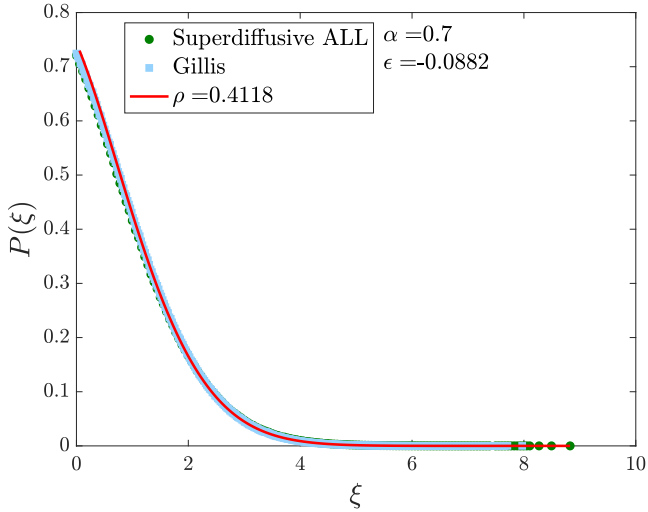}%
	}
	\caption{Distributions of the occupation time (top) and the number of returns (bottom) for the averaged L\'evy-Lorentz gas, superdiffusive version (green points), and the Gillis random walk (light blue squares), compared to the theoretic result. The values of the corresponding parameters are $ \alpha=0.7 $ and $ \epsilon=-0.0882 $, which in both cases yield $ \rho= 7/17$. For both systems we considered $ 10^7 $ walks evolved for $ 10^4 $ steps.}
	\label{fig:Gill_vs_LL_rho=7/17}
\end{figure}

\begin{figure}
	\subfloat{
		\includegraphics[width=8.5cm]{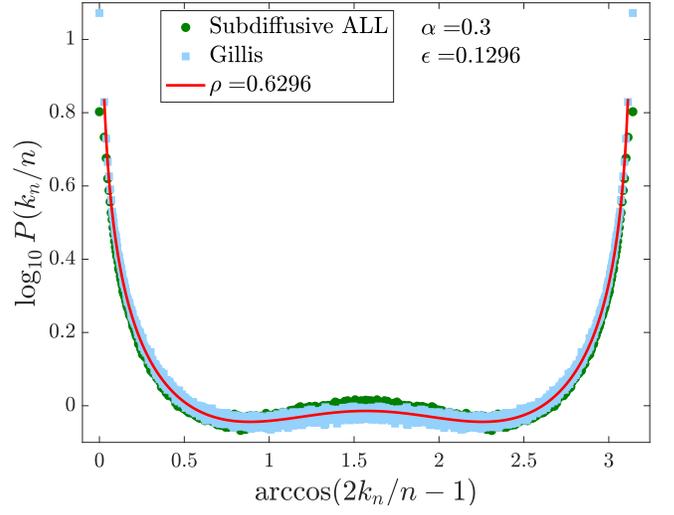}%
	}\hfill
	\subfloat{
		\includegraphics[width=8.5cm]{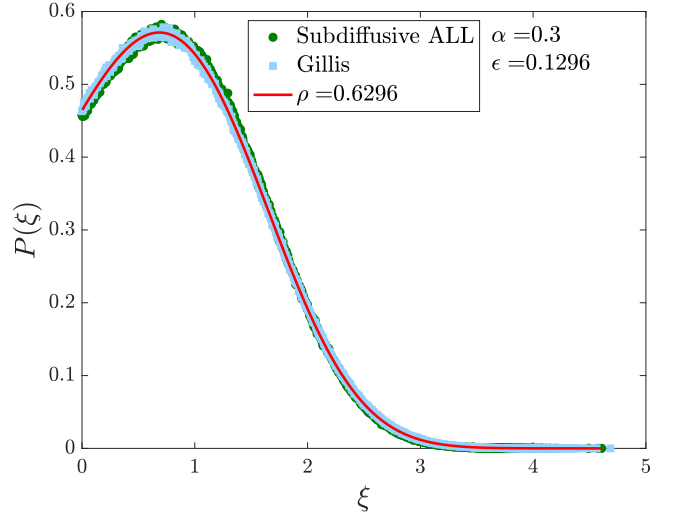}%
	}
	\caption{Distributions of the occupation time (top) and the number of returns (bottom) for the averaged L\'evy-Lorentz gas, subdiffusive version (green points), and the Gillis random walk (light blue squares), compared to the theoretic result. The values of the corresponding parameters are $ \alpha=0.3 $ and $ \epsilon=0.1296 $, which in both cases yield $ \rho= 17/27$. For both systems we considered $ 10^7 $ walks evolved for $ 10^4 $ steps.}
	\label{fig:Gill_vs_LL_rho=17/27}
\end{figure}

In order to show this, we compare the two distributions for the Gillis random walk and both versions of the averaged L\'evy-Lorentz gas. For the latter system we consider the values of $ \alpha $ already chosen in the previous sections, viz. $ \alpha=0.7 $ for the superdiffusive version and $ \alpha=0.3 $ for the subdiffusive one. The two corresponding values of the Lamperti parameter are $ \rho=\frac 7 {17} $ (superdiffusive) and $ \rho=\tfrac {17}{27} $ (subdiffusive), which are obtained in the case of the Gillis random walk for $ \epsilon=-0.0882 $ and $ \epsilon=0.1296 $, respectively. The results are presented in Figs. \ref{fig:Gill_vs_LL_rho=7/17} and \ref{fig:Gill_vs_LL_rho=17/27}. In both cases the simulations agree with the theoretical predictions.

\section{Conclusions and discussion}\label{s:Concl}
We have shown that for a general class of stochastic processes there is a deep connection between the statistics of the occupation times, the number of visits at the origin and the survival probability. The distributions of these observables are all characterized by a single parameter, which is related to the asymptotic power-law decay of the probability of occupying the origin.

We point out that the results of this paper are also associated with infinite ergodic theory. In particular, let us consider the Darling-Kac theorem, that we used in Sec. \ref{s:DK} to obtain the statistics of the occupation time of the origin, in its continuous-time version \cite{Dar-Kac}. The theorem firstly requires that, for a given non-negative and integrable function $ V(x) $, one has
\begin{equation}\label{eq:concl_DK_Hyp}
\lim_{s\to 0} \frac{1}{\pi(s)}\int P_s(x|x_0)V(x)\mathrm{d}x=c,
\end{equation}
where $ c $ is a positive constant, $ P_s(x|x_0) $ is the Laplace transform from $ t $ to $ s $ of the probability of arriving at $ x $ starting from $ x_0 $ in time $ t $, and $ \pi(s) $ is a function such that $ \pi(s)\to\infty $ as $ s\to 0 $. Now suppose that we have
\begin{equation}\label{eq:concl_Infinite_density}
\lim_{s\to 0} \frac{P_s(x|x_0)}{\pi(s)}=\mathcal{I}_\infty(x),
\end{equation}
where $ \mathcal{I}_\infty(x) $ is called, in the language of infinite ergodic theory, the infinite density \cite{Lei-Bar}, since it is not normalizable. Note that in this case, if $ V(x) $ is measurable with respect to the infinite density, the condition given in Eq. \eqref{eq:concl_DK_Hyp} is satisfied. Therefore, if $ \pi(s)=s^{-\rho}H(1/s) $, with $ H(u) $ slowly-varying, the Darling-Kac theorem states that the random variable
\begin{align}\label{key}
\xi &=\lim_{t\to\infty}\frac 1{c\pi(1/t)}\int_{0}^{t}V(x(\tau))\mathrm{d}\tau\\
	&=\lim_{t\to\infty}\frac 1{ct^\rho H(t)}\int_{0}^{t}V(x(\tau))\mathrm{d}\tau
\end{align}
follows a Mittag-Leffler distribution of order $ \rho $. Now we observe that using Eq. \eqref{eq:concl_Infinite_density} we can say
\begin{equation}\label{key}
P_s(x|x_0)\sim \pi(s)\mathcal{I}_\infty(x)
\end{equation}
and therefore for the ensemble average of $ V(x) $ we have
\begin{equation}\label{key}
\langle V_s\rangle=\int P_s(x|x_0)V(x)\mathrm{d}x\sim c\pi(s).
\end{equation}
In the case $ \pi(s)=s^{-\rho}H(1/s) $, by using the tauberian theorem \cite{Fell-I} we find that the ensemble average in the long-time limit behaves as
\begin{equation}\label{key}
\langle V_t\rangle\sim \frac c{\Gamma(\rho)}t^{\rho-1}H(t)
\end{equation}
and therefore
\begin{equation}\label{key}
\xi=\lim_{t\to\infty}\frac{1}{\Gamma(\rho)}\frac{\overline{V}_t}{\langle V_t\rangle},
\end{equation}
where $ \overline{V}_t $ indicates the time average of $ V(x) $ over a single realization. Such a ratio is a random variable distributed according to a Mittag-Leffler of order $ \rho $. This is the main difference with standard ergodic theory, where instead time averages converge to ensemble averages, and hence $ \xi $ is expected to be distributed according to a Dirac delta function centered around $ \xi_0=1 $. Now the important point is that the scaling function $ \pi(s) $, which determines the distribution of $ \xi $, is a property of the propagator $ P_s(x|x_0) $. In other words, for any function which is measurable with respect to the infinite density, the distribution of $ \xi $ only depends on the long-time properties of the propagator. Therefore, it is possible to determine the distribution of $ \xi $ by just evaluating the long-time behavior of the probability distribution in a given set, as we have done in the paper by considering the probability of occupying the origin. However, in general it is not possible to formulate the connection between the Lamperti distribution, the Mittag-Leffler distribution and the survival probability, since if we focus on a point off the origin, we are not able to use the simple relation, Eq. \eqref{eq:F_vs_P}, between the first return probability and the occupation probability.

\begin{acknowledgments}
The authors acknowledge partial support by the research project PRIN 2017S35EHN\_007 ``Regular and stochastic behaviour in dynamical systems'' of the Italian Ministry of Education and Research.
\end{acknowledgments}

\appendix
\section{Meaning of the random variable $ T_n $}\label{ap:Tn}
We consider the random variable
\begin{equation}\label{key}
T_n\equiv\frac{1}{H(n)n^\rho}\sum_{m=0}^n\delta_{x_m,0}.
\end{equation}
The sum
\begin{equation}\label{key}
M_n=\sum_{m=0}^{n}\delta_{x_m,0}
\end{equation}
clearly represents the number of times the random walk has visited the origin up to time $ n $, while it is possible to show that the denominator $ n^\rho H(n) $ is connected to the asymptotics of the mean occupation time. Indeed, for $ M\geq 1 $, let us call $ \psi_n(M) $ the probability that the $ M $-th visit occurs at step $ n $, and $ U_n $ the probability of observing no returns to the origin up to step $ n $, with the initial conditions $ \psi_0(M)=\delta_{M,1} $ and $ U_0=1 $. We have
\begin{align}
U_n &= 1-\sum_{m=0}^{n}F_m \label{eq:U_n} \\
\psi_n(1) &=\delta_{n,0} \label{eq:ML_phi_1},
\end{align}
while for $ M\geq 2 $ we can write the recurrence relation
\begin{equation}\label{eq:ML_phi_M_recurr}
\psi_n(M)=\sum_{m=0}^{n}F_m\psi_{n-m}(M-1).
\end{equation}
From equations \eqref{eq:U_n}, \eqref{eq:ML_phi_1} and \eqref{eq:ML_phi_M_recurr} we can compute the generating functions
\begin{align}
U(z) &= \frac{1-F(z)}{1-z}\label{eq:U(z)}\\
\psi_z(M) &= \left[F(z)\right]^{M-1}.
\end{align}
Now, the probability $ \phi_n(M) $ of $ M $ visits in $ n $ steps is equal to the probability that the $ M $-th visit has occurred at step $ k\leq n $, and then no other visit occurs up to time $ n $:
\begin{equation}\label{key}
\phi_n(M)=\sum_{m=0}^{n}\psi_k(M)U_{n-m},
\end{equation}
hence its generating function reads
\begin{equation}\label{key}
\phi_z(M)=F^{M-1}(z)\frac{1-F(z)}{1-z}.
\end{equation}
The generating function of the mean number of visits is
\begin{equation}\label{key}
\langle M(z)\rangle = \sum_{M=1}^{\infty}M\phi_z(M)=\frac{1}{1-z}\frac{1}{1-F(z)}
\end{equation}
and since we know the relation between $ F(z) $ and $ P(z) $, Eq. \eqref{eq:F_vs_P}, and the form that $ P(z) $ must assume, Eq. \eqref{eq:P(z)_form}, we have
\begin{equation}\label{key}
\langle M(z)\rangle = \frac{1}{(1-z)^{1+\rho}}H\left(\frac{1}{1-z}\right),
\end{equation}
and the tauberian theorem implies:
\begin{equation}\label{key}
\langle M_n\rangle \sim \frac{1}{\Gamma(1+\rho)}n^\rho H(n),
\end{equation}
which is valid for $ 0\leq\rho\leq 1 $. We conclude that the random variable $ T_n $ represents, up to a constant factor, the asymptotic value of the occupation time of the origin rescaled for its mean value:
\begin{equation}\label{key}
T_n\sim\frac{1}{\Gamma(1+\rho)}\frac{M_n}{\langle M_n\rangle}.
\end{equation}

\section{The relation between the survival and persistence probabilities and their asymptotic behavior}\label{ap:Q_vs_U}
We consider the survival probability in the set $ A $. Define
\begin{align}
F_n &= \mathrm{Pr}\lbrace x_1\neq 0, x_2\neq 0,\ldots,x_n=0|x_0=0\rbrace\\
Q_n &= \mathrm{Pr}\lbrace x_1\geq 0, x_2\geq 0,\ldots,x_n\geq 0|x_0=0\rbrace\\
U_n &= \mathrm{Pr}\lbrace x_1\neq 0, x_2\neq 0,\ldots,x_n\neq 0|x_0=0\rbrace,
\end{align}
with the initial conditions $ F_0=0$, $Q_0=1$ and $U_0=1 $, and the generating functions
\begin{align}
F(z) &= \sum_{n=1}^{\infty}F_nz^n\\
Q(z) &= \sum_{n=0}^{\infty}Q_nz^n\\
U(z) &= \sum_{n=0}^{\infty}U_nz^n.
\end{align}
 It is easy to see that if the process is symmetric with respect to the two sets, the following relation holds:
\begin{equation}\label{key}
2Q_n=\delta_{n,0}+U_n+\sum_{m=1}^{n}F_mQ_{n-m}.
\end{equation}
By passing to the generating function we get
\begin{equation}\label{key}
2Q(z)=1+U(z)+F(z)Q(z)
\end{equation}
and by using Eq. \eqref{eq:U(z)} in appendix \ref{ap:Tn}, after some algebra we obtain
\begin{equation}\label{eq:ap_Q_vs_U}
Q(z)=\frac{1+U(z)}{1+\left(1-z\right)U(z)}.
\end{equation}

To show that $ Q(z) $ and $ U(z) $ have the same $ z\to 1 $ behavior, we use a result by Karamata \cite{Kar-1962}: if $ L(x) $ is a slowly varying function, then for any $ \gamma>0 $:
\begin{align}\label{key}
\lim_{x\to \infty}x^{-\gamma}L(x)=0\\
\lim_{x\to \infty}x^{\gamma}L(x)=\infty.
\end{align}
We showed in the main text, Eq. \eqref{eq:U(z)_sv}, that $ U(z) $ is of the form:
\begin{equation}\label{key}
U(z)=\frac 1{(1-z)^{1-\rho}}L\left(\frac{1}{1-z}\right),
\end{equation}
therefore, as $ z\to 1 $, $ U(z) $ diverges and $ (1-z)U(z) $ converges to $ 0 $. For $ \rho=0 $ we still have the divergence of $ U(z) $, but we cannot use the previous result by Karamata for $ (1-z)U(z) $, because
\begin{equation}\label{key}
(1-z)U(z)=L\left(\frac{1}{1-z}\right).
\end{equation}
However, since in this case
\begin{equation}\label{key}
F(z)=1-L\left(\frac{1}{1-z}\right)
\end{equation}
and recurrence implies $ F(z)\to 1 $, we still have $ (1-z)U(z)\to 0 $.  Hence, it follows from Eq. \eqref{eq:ap_Q_vs_U} that $ Q(z)\sim U(z) $ for any $ 0\leq\rho<1 $.

\section{Evaluation of the Lamperti parameter for the Gillis random walk}\label{ap:rho_Gillis}
The strategy is to put the generating function $ P(z) $, Eq. \eqref{eq:Gillis_genfun} in the main text, in the form:
\begin{equation}\label{key}
P(z)=\frac{1}{(1-z)^{\nu}}H\left(\frac{1}{1-z}\right),
\end{equation}
where $ H(x) $ is a slowly-varying function. We make use of the transformation formulae \cite{Abr-Steg}:
\begin{widetext}
\begin{multline}\label{eq:hyp_trans_nonint}
\hyp(a,b;c;z)=\,\frac{\Gamma(c)\Gamma(c-a-b)}{\Gamma(c-a)\Gamma(c-b)}\,\hyp(a,b;a+b-c+1;1-z) \\
+(1-z)^{c-a-b}\frac{\Gamma(c)\Gamma(a+b-c)}{\Gamma(a)\Gamma(b)}\,\hyp(c-a,c-b;c-a-b+1;1-z),
\end{multline}
valid for $ c-a-b $ non-integer, while for the integer case we use
\begin{multline}
\hyp(a,b;a+b+m;z)=\,\frac{\Gamma(m)\Gamma(a+b+m)}{\Gamma(a+m)\Gamma(b+m)}\sum_{n=0}^{m-1}\frac{(a)_n(b)_n}{n!(1-m)_n}(1-z)^n\\
-(z-1)^{m}\frac{\Gamma(a+b+m)}{\Gamma(a)\Gamma(b)}\sum_{n=0}^{\infty}\frac{(a+m)_n(b+m)_n}{n!(n+m)!}(1-z)^n\times\\
\times\left[\log(1-z)-\psi(n+1)-\psi(n+m+1)+\psi(a+n+m)+\psi(b+n+m)\right]
\end{multline}
and
\begin{multline}\label{key}
\hyp(a,b;a+b-m;z)=\,(1-z)^{-m}\frac{\Gamma(m)\Gamma(a+b-m)}{\Gamma(a)\Gamma(b)}\sum_{n=0}^{m-1}\frac{(a-m)_n(b-m)_n}{n!(1-m)_n}(1-z)^n\\
-(-1)^{m}\frac{\Gamma(a+b-m)}{\Gamma(a-m)\Gamma(b-m)}\sum_{n=0}^{\infty}\frac{(a)_n(b)_n}{n!(n+m)!}(1-z)^n\times\\
\times\left[\log(1-z)-\psi(n+1)-\psi(n+m+1)+\psi(a+n)+\psi(b+n)\right]
\end{multline}
for $ m=1,2,\dots $, or
\begin{equation}\label{key}
\hyp(a,b;a+b,z)=\frac{\Gamma(a+b)}{\Gamma(a)\Gamma(b)}\sum_{n=0}^{\infty}\frac{(a)_n(b)_n}{\left(n!\right)^2}(1-z)^n\left[2\psi(n+1)-\psi(a+n)-\psi(b+n)-\log(1-z)\right],
\end{equation}
\end{widetext}
where $ \psi(z)\equiv\tfrac{\mathrm{d}}{\mathrm{d}z}\log\Gamma(z) $ is the digamma function, and $ \left( z\right)_n \equiv \Gamma\left(z+n\right)/\Gamma(z)$ denotes the Pochhammer's symbol \cite{Abr-Steg}.

Now, since the generating function $ P(z) $
\begin{equation}\label{key}
P(z)=\frac{\hyp\left(\tfrac{1}{2}\epsilon+1,\tfrac{1}{2}\epsilon+\tfrac{1}{2};1;z^2\right)}{\hyp\left(\tfrac{1}{2}\epsilon,\tfrac{1}{2}\epsilon+\tfrac{1}{2};1;z^2\right)}
\end{equation}
is a function of $ z^2 $, for the sake of simplicity we consider
\begin{align}\label{eq:ap_Gil_genfun_sqrt}
P\left(\sqrt{z}\right)\equiv\Pi(z)&=\sum_{n=0}^\infty\varpi_nz^n\nonumber\\
&=\frac{\hyp\left(\tfrac{1}{2}\epsilon+1,\tfrac{1}{2}\epsilon+\tfrac{1}{2};1;z\right)}{\hyp\left(\tfrac{1}{2}\epsilon,\tfrac{1}{2}\epsilon+\tfrac{1}{2};1;z\right)},
\end{align}
so that the $ n $-th coefficient $ \varpi_n $ corresponds to $ P_{2n} $. It is easy to show that if $ P(z) $ is of the form
\begin{equation}\label{key}
P(z)=\frac{1}{(1-z)^\rho}H\left(\frac{1}{1-z}\right),
\end{equation}
then also $ \Pi(z) $ can be written as
\begin{equation}\label{key}
\Pi(z)=\frac{1}{(1-z)^\rho}G\left(\frac{1}{1-z}\right),
\end{equation}
where $ G(x) $ is slowly-varying and related to $ H(x) $ by
\begin{equation}\label{eq:ap_Gil_G_vs_H}
G(x)=\frac{1}{x^\rho\left(1-\sqrt{1-\tfrac 1x}\right)^\rho}H\left(\frac 1{1-\sqrt{1-\tfrac 1x}}\right).
\end{equation}
This means that the transformation does not change the exponent $ \rho $. By using Eq. \eqref{eq:ap_Gil_genfun_sqrt} we obtain the following results:

\begin{widetext}
\begin{itemize}	
\item[1)]In the case $ \epsilon = -\tfrac 12 $ we get
	\begin{equation}\label{key}
	\Pi(z)=G\left(\frac{1}{1-z}\right),
	\end{equation}
	where the slowly-varying function is
	\begin{equation}\label{key}
	G\left(x\right)=\frac{\sum_{n=0}^{\infty}\tfrac{(3/4)_n(1/4)_n}{\left(n!\right)^2}(x)^{-n}\left[2\psi(n+1)-\psi\left(\tfrac 34+n\right)-\psi\left(\tfrac 14+n\right)+\log(x)\right]}{4+\tfrac{1}{4}\sum_{n=0}^{\infty}\tfrac{(3/4)_n(5/4)_n}{n!(n+1)!}(x)^{-n-1}\left[\log(x)+\psi(n+1)+\psi(n+2)-\psi\left(\tfrac 34 + n\right)-\psi\left(\tfrac 54+n\right)\right]};
	\end{equation}
	
\item[2)]In the range $ \epsilon \in \left(-\tfrac 12,\tfrac 12\right) $ the generating function has the form
	\begin{equation}\label{key}
	\Pi(z)=\frac{1}{\left(1-z\right)^{1/2 +\epsilon}}G\left(\frac{1}{1-z}\right)
	\end{equation}
	with
	\begin{equation}\label{key}
	G(x)=a_1\frac{\,_2F_1\left(-\tfrac{1}{2}\epsilon,\tfrac{1}{2}-\tfrac{1}{2}\epsilon;\tfrac 12-\epsilon;\tfrac{1}{x}\right)+a_2x^{-1/2-\epsilon}\,_2F_1\left(\tfrac{1}{2}\epsilon+1,\tfrac{1}{2}\epsilon+\tfrac{1}{2};\tfrac{3}{2}+\epsilon;\tfrac 1x\right)}{\,_2F_1\left(\tfrac{1}{2}\epsilon,\tfrac{1}{2}\epsilon+\tfrac{1}{2};\tfrac 12+\epsilon;\tfrac{1}{x}\right)+a_3x^{-1/2+\epsilon}\,_2F_1\left(1-\tfrac{1}{2}\epsilon,\tfrac{1}{2}-\tfrac{1}{2}\epsilon;\tfrac{3}{2}-\epsilon;\tfrac 1x\right)},
	\end{equation}
	where $ a_1, a_2 $ and $ a_3 $ are numerical coefficients (depending on $ \epsilon $) which can be determined from formula \eqref{eq:hyp_trans_nonint};
	
\item[3)]For $ \epsilon=\tfrac 12 $ we have
	\begin{equation}\label{key}
	\Pi(z)=\frac{1}{1-z}G\left(\frac{1}{1-z}\right)
	\end{equation}
	where $ G(x) $ has the expression
	\begin{equation}\label{key}
	G\left(x\right)=\frac{4-\tfrac{1}{4}\sum_{n=0}^{\infty}\tfrac{(5/4)_n(3/4)_n}{n!(n+1)!}(x)^{-n-1}\left[\log(x)+\psi(n+1)+\psi(n+2)-\psi\left(\tfrac 54 + n\right)-\psi\left(\tfrac 34+n\right)\right]}{\sum_{n=0}^{\infty}\tfrac{(1/4)_n(3/4)_n}{\left(n!\right)^2}(x)^{-n}\left[2\psi(n+1)-\psi\left(\tfrac 14+n\right)-\psi\left(\tfrac 34+n\right)+\log(x)\right]};
	\end{equation}
	
\item[4)]Finally when $ \epsilon\in\left(\tfrac 12,1\right) $ the generating function has the same form as the previous case,
	\begin{equation}\label{key}
	\Pi(z)=\frac{1}{1-z}G\left(\frac{1}{1-z}\right),
	\end{equation}
	but with
	\begin{equation}\label{key}
	G(x)=b_1\frac{\,_2F_1\left(-\tfrac{1}{2}\epsilon,\tfrac{1}{2}-\tfrac{1}{2}\epsilon;\tfrac 12-\epsilon;\tfrac{1}{x}\right)+b_2x^{-1/2-\epsilon}\,_2F_1\left(\tfrac{1}{2}\epsilon+1,\tfrac{1}{2}\epsilon+\tfrac{1}{2};\tfrac{3}{2}+\epsilon;\tfrac 1x\right)}{\,_2F_1\left(1-\tfrac{1}{2}\epsilon,\tfrac{1}{2}-\tfrac{1}{2}\epsilon;\tfrac 32-\epsilon;\tfrac{1}{x}\right)+b_3x^{1/2-\epsilon}\,_2F_1\left(\tfrac{1}{2}\epsilon,\tfrac{1}{2}\epsilon+\tfrac{1}{2};\tfrac{1}{2}+\epsilon;\tfrac 1x\right)}
	\end{equation}
	where once again $ b_1 $, $ b_2 $ and $b_3 $ can be determined from Eq. \eqref{eq:hyp_trans_nonint}.
\end{itemize}
\end{widetext}
We remark that if one wishes to go back to $ P(z) $, it is now sufficient to recover the expression of $ H(x) $ by using Eq. \eqref{eq:ap_Gil_G_vs_H}.

\bibliographystyle{apsrev4-2}
\end{document}